\documentclass[letterpaper, 10 pt, conference]{ieeeconf}  

\IEEEoverridecommandlockouts                              

\usepackage{graphics} 
\usepackage{amsmath} 
\usepackage{amssymb}  
\usepackage{mathtools}

\usepackage{amsthm}
\usepackage{bbold}
\usepackage{cite}
\usepackage{subcaption}
\usepackage{url}
\usepackage{xcolor}

\title{\LARGE \bf
A Continuification-Based Control Solution for Large-Scale Shepherding
}

\author{Beniamino Di Lorenzo$^{1}$, Gian Carlo Maffettone$^{1}$ and Mario di Bernardo$^{1, 2,*}$
\thanks{This work was supported financially by the Italian Ministry of University and Research (MUR) under the PRIN 2022 project titled 'Machine-learning-based control of complex multi-agent systems for search and rescue operations in natural disasters (MENTOR).}
\thanks{$^{1}$Modeling and Engineering Risk and Complexity, Scuola Superiore Meridionale, Largo san Marcellino 10, 80136, Naples, Italy. (emails:
        {ben.dilorenzo@studenti.unina.it, giancarlo.maffettone@unina.it, mario.dibernardo@unina.it})}%
\thanks{$^{2}$Department of Electrical Engineering and Information Technology, Univeristy of Naples Federico II, via Claudio 21, 80125, Naples, Italy.}%
\thanks{$^*$ corresponding author}
}

\newtheorem{theorem}{Theorem}
\newtheorem{definition}{Definition}

\newtheorem{proposition}{Proposition}

\newcommand{\xt}{\left(\mathbf{x},t\right)}

\begin{document}

\maketitle
\thispagestyle{empty}
\pagestyle{empty}

\begin{abstract}
In this paper, we address the large-scale shepherding control problem using a continuification-based strategy. We consider a scenario in which a large group of follower agents (targets) must be confined within a designated goal region through indirect interactions with a controllable set of leader agents (herders). Our approach transforms the microscopic agent-based dynamics into a macroscopic continuum model via partial differential equations (PDEs). This formulation enables efficient, scalable control design for the herders' behavior, with guarantees of global convergence. Numerical and experimental validations in a mixed-reality swarm robotics framework demonstrate the method’s effectiveness.
\end{abstract}

\section{INTRODUCTION} \label{sec:introduction}


Many new methodologies for controlling large-scale multi-agent systems rely on their macroscopic descriptions, typically represented by partial differential equations (PDEs), capturing a collective behavior of interest. For instance, describing the spatio-temporal dynamics of the density of a swarming group is often more practical than modeling each agent individually, inherently requiring a prohibitively large set of ordinary or stochastic differential equations (ODEs/SDEs) for velocities and accelerations \cite{maffettone2022continuification, maffettone2024leaderfollower, nikitin2021continuation, bongini2017optimal, fornasier2014mean}. Hence, macroscopic approaches offer several advantages, including compactness and analytical tractability, making the mathematical formulation scalable and able to address the curse of dimensionality \cite{d2023controlling}. 

These macroscopic solutions are typically derived under the assumption of an infinite number of agents, presenting a pressing challenge in assessing their applicability to real-world applications. In this paper, we focus on a large-scale version of the shepherding control problem \cite{lama2024shepherding, long2020comprehensive}. Inspired by the shepherding behavior exhibited by sheepdogs, we consider a scenario where a large group of follower agents (targets) must be guided into a desired region through interactions with a population of leader agents (herders) \cite{jyh2004shepherding}. This control problem has practical applications, such as guiding robotic systems for environmental pollutant containment \cite{zahugi2013oil} or directing biological agents (e.g. animal or insects) to safer regions in search and rescue operations \cite{yuan2023multi}.

\begin{figure}
    \centering
    \includegraphics[width=1\linewidth]{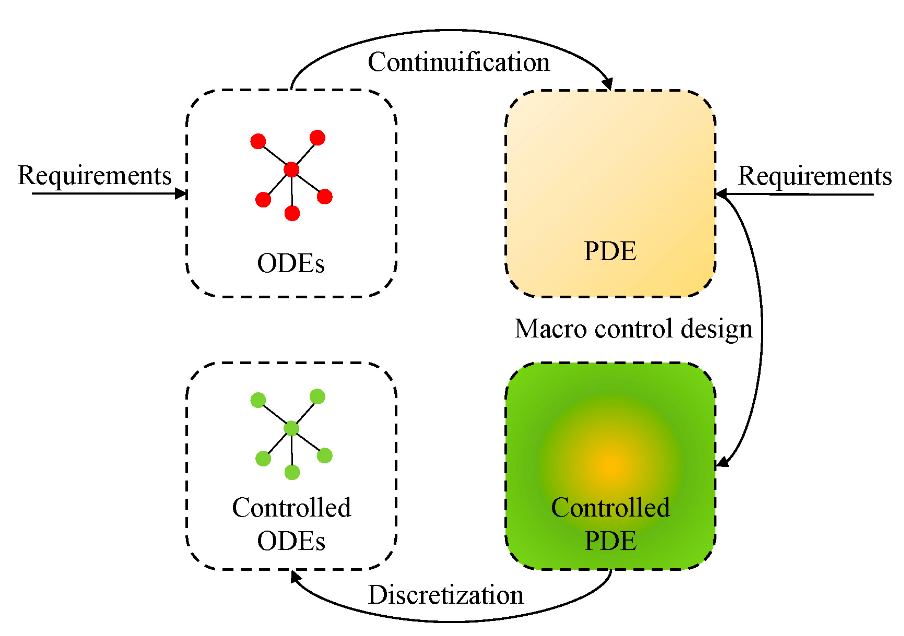}
    \caption{Continuification control pipeline, inspired by \cite{nikitin2021continuation}.}
    \label{fig:continuification_pipeline}
\end{figure}

Addressing this control problem using traditional microscopic techniques becomes challenging in large-scale scenarios. In many large-scale applications, including the shepherding control problem, the objective is to influence the system’s macroscopic behavior, yet control can only be applied at the microscopic agent level \cite{d2023controlling}. A feasible macroscopic approach to these multi-scale control problems is continuification-based control \cite{nikitin2021continuation, maffettone2022continuification}. This method derives a continuum model from the agent-based dynamics, represented as ODEs/SDEs, and reformulates it into a compact set of PDEs to facilitate control law design at the macroscopic level. The resulting continuum control law is ultimately discretized to generate the microscopic inputs required at the agent level (see Fig. \ref{fig:continuification_pipeline} for a schematic).

In this work, we leverage the continuification control pipeline and the theoretical foundations established in \cite{maffettone2024leaderfollower} to address open challenges in large-scale shepherding control. Specifically, we ($i$) determine the minimum number of herders and their required sensing capabilities and ($ii$) guide the targets to a desired goal region, providing analytical guarantees of global convergence in the continuum limit.

We validate the proposed strategy experimentally using a mixed-reality platform with 20 differential-drive robots and a configurable number of virtual agents, similar to the set-up presented in \cite{maffettone2024mixed}. This experimental platform was developed utilizing the facilities provided by the Robotarium \cite{robotarium}.


\section{MODEL AND PROBLEM STATEMENT} \label{sec:model}
\subsection{The model}
We consider an interacting group of $N^H$ herders and $N^T$ targets moving on the periodic square $\Omega = [-\pi, \pi]^2$. Under the kinematic assumption (neglecting
acceleration and considering a drag force proportional to the
velocity), similarly to \cite{lama2024shepherding}, we model herders as deterministic single integrators and targets as stochastic, non-cohesive, agents, yielding the following set of ODEs/SDEs,
\begin{subequations}\label{eq:agents}
\begin{align}
    \dot{\mathbf{H}}_i &= \mathbf{u}_i, \label{eq:herders_discrete}\\
    \mathrm{d}\mathbf{T}_k &= \alpha \sum_{j=1}^{N^H} \mathbf{f} \left(\{ \mathbf{T}_k,\mathbf{H}_j \}\right)\mathrm{d}t + \sqrt{2D} \mathrm{d}\mathbf{B}_k,  \label{eq:targets_discrete}
\end{align}
\end{subequations}
where  $\mathbf{H}_i \in \Omega$, $i=1,\dots,N^H$ and $\mathbf{T}_k \in \Omega$, $k = 1,\dots,N^T$, are the positions of herders and targets respectively, $\mathbf{u}_i$ is a velocity control input steering the herders' behavior, $\alpha$ is a normalization factor that will be used later to normalize the total mass of agents to unity in the continuum framework, and $\mathbf{f}:\Omega \rightarrow \mathbb{R}^2$ is an odd periodic soft-core velocity interaction kernel capturing the influence of herders on targets (e.g., by repulsion, by attraction, or by a combination of attraction and repulsion at different ranges, see \cite{maffettone2022continuification} for further details). Moreover, $\{ \mathbf{T}_k,\mathbf{H}_j \}$ is the relative position between target $k$ and herder $j$ wrapped to have values in $\Omega$,
and $\mathbf{B}_k$ is a standard Wiener process weighted by the diffusion coefficient $D\in\mathbb{R}_{>0}$.

As often done in the shepherding literature (e.g., \cite{ko2020asymptotic,lama2024shepherding}), we consider the interaction kernel $\mathbf{f}$ to be repulsive. Specifically, we recover the periodic repulsive kernel from its non-periodic counterpart
\begin{equation} \label{eq:repulsive_kernel}
    \hat{\mathbf{f}}(\mathbf{x}) = \mathrm{\textbf{sign}}(\mathbf{x}) \, \mathrm{e}^{-\frac{\| \mathbf{x} \|_2}{L}}
\end{equation}
where $\mathrm{\textbf{sign}}(\mathbf{x}) = \mathbf{x}/\| \mathbf{x} \|_2$, and $L$ is the characteristic interaction length (see \cite{maffettone2024mixed} and \cite{boldini2024stigmergy} for more details).

Assuming a sufficiently large number of agents, we can describe the system’s spatial organization compactly through the density functions of herders and targets, denoted by $\rho^H, \rho^T: \Omega \times \mathbb{R}_{\geq 0} \rightarrow \mathbb{R}_{\geq 0}$. When integrated over the domain $\Omega$, these functions yield the total mass of herders and targets, $M^H$ and $M^T$.
Additionally,  we fix $\alpha = 1/(N^H+N^T)$ in \eqref{eq:targets_discrete} to enforce the constraint $M^T + M^H = 1$.

\subsection{Problem statement}
Given the interaction kernel length $L$, the diffusion coefficient $D$ characterizing the targets’ dynamics, and the number of targets $N^T$, the objective is to determine the appropriate number of herders $N^H$ and the distributed control inputs $\mathbf{u}_i$ at the microscopic level able to drive the herders’ movements to ensure that the targets enter asymptotically some desired goal region in the plane.  
For simplicity, we assume that the goal region is a circular area with radius $r^*$, centered at a point $\mathbf{p} \in \Omega$. The control objective is thus formulated as achieving the condition
\begin{equation} \label{eq:problem_statement_herding}
    \lim_{t\to\infty} \left\| \mathbf{T}_i(t) - \mathbf{p} \right\|_2 \leq r^*, \quad i = 1,\dots,N^F,
\end{equation}
for agents (herders and targets) starting from any initial configuration in the plane.

\section{CONTROL DESIGN} \label{sec:control_design}
We solve the problem using the continuification-based approach proposed in \cite{maffettone2022continuification}. In a large-scale scenario, problem \eqref{eq:problem_statement_herding} can be reframed as a leader-follower density control problem, as that in \cite{maffettone2024leaderfollower}. From this perspective, the task of confining targets is transformed into guiding the targets' density $\rho^T$ toward a desired profile $\bar{\rho}^T$, which is selected to match the target goal region.  Specifically, the problem is reformulated as finding a set of distributed control inputs $\mathbf{u}_i$, for $i = 1, \dots, N^H$ in \eqref{eq:herders_discrete} such that
\begin{equation} \label{eq:continuum_problem_statement}
    \lim_{t\to\infty} \left\| \Bar{\rho}^T(\cdot) - \rho^T(\cdot,t) \right\|_2 = 0
\end{equation}

To achieve confinement within the circular region specified in \eqref{eq:problem_statement_herding}, we select the desired targets' density to follow a Von Mises distribution given by
\begin{equation} \label{eq:von_mises}
    \Bar{\rho}^T(\mathbf{x}) = Z\exp\{ \mathbf{k}^\intercal \mathbf{c}_1(\mathbf{x},\mu,\nu) + \mathbf{c}_2(\mathbf{x},\mu,\mu) \mathbf{I}_2 \mathbf{c}_2^\intercal (\mathbf{x},\nu,\nu) \},
\end{equation}
with $\mathbf{k} = [k_1, k_2]^\intercal$ being the vector of concentration coefficients, $\mu$ and $\nu$ the means along each direction, $\mathbf{c}_1=[\cos(x_1-a), \cos(x_2-b)]$, $\mathbf{c}_2 = [\cos(x_1-a), \sin(x_2-b)]$ ($a,b\in\Omega$), where $x_1,x_2$ are the components of $\mathbf{x}$ in the Cartesian coordinate system, and $\mathbf{I}_2$ is the second-order identity matrix. $Z$ is a normalization coefficient to allow $\Bar{\rho}^T$ to sum to the total targets' mass. In order to match the geometry of the desired goal region, we set $[\mu, \nu]^\intercal = \mathbf{p}$, and $k_1 = k_2 = 3/r^*$. This involves centering the Von Mises distribution at $\mathbf{p}$ and setting its concentration parameters so that the density is nearly zero outside the circle of radius $r^*$.

We next outline the solution to the density control problem through the three stages of the continuification pipeline: continuification, macroscopic control design, and discretization.

\subsection{Continuification}\label{sec:continuification}
We reformulate the microscopic agents’ dynamics using a continuum approach, as proposed in \cite{maffettone2024leaderfollower}. The herders' dynamics is described by a mass conservation law, which we can fully actuate towards its characteristic velocity field. The targets' dynamics is modeled by a convection-diffusion equation, where the convection term captures the influence of herders on targets' movement, and the diffusion term accounts for the microscopic stochastic behavior. 

This yields the following continuum model:
\begin{subequations} \label{eq:LF_continuum_model}
    \begin{align}
        \begin{split} \label{eq:herders_continuum}
            \rho_t^H(\mathbf{x},t) + \nabla \cdot \left[ \rho^H(\mathbf{x},t)\mathbf{u}(\mathbf{x},t) \right] &= 0 ,
        \end{split} \\
        \begin{split} \label{eq:targets_continuum}
            \rho_t^T(\mathbf{x},t) + \nabla \cdot \left[ \rho^T(\mathbf{x},t)\mathbf{v}^{TH}(\mathbf{x},t) \right] &= D\nabla^2\rho^T(\mathbf{x},t),
        \end{split}
    \end{align}
\end{subequations}
where $\mathbf{x}\in\Omega$ and $t\in\mathbb{R}_{\geq0}$ are the space and time variables, and $\mathbf{u} : \Omega \times \mathbb{R}_{\geq 0} \rightarrow \mathbb{R}^2$ is the velocity field to be designed to control the herders' dynamics. The cross-convectional term in \eqref{eq:targets_continuum} captures the influence of herders on targets through the velocity field 
\begin{equation} \label{eq:v_FL}
    \mathbf{v}^{TH}\xt = \int_\Omega \mathbf{f}\left(\{\mathbf{x},\mathbf{y}\}\right) \rho^H(\mathbf{y},t) \, \mathrm{d} \mathbf{y} = \left( \mathbf{f} * \rho^H \right)\xt,
\end{equation}
where ``$*$" is the two-dimensional circular convolution operator. 
For \eqref{eq:herders_continuum} to be well posed, we require periodicity of $\mathbf{u}$ and $\rho^H$ on $\partial \Omega$ for all $t\in \mathbb{R}_{\geq 0}$. Indeed, integrating \eqref{eq:herders_continuum} and exploiting the divergence theorem and the periodicity of the flux, we get $\left(\int_\Omega \rho^H \xt \, \mathrm{d} \mathbf{x} \right)_t = 0$. As for \eqref{eq:targets_continuum}, being $\mathbf{v}^{TH}$ periodic by construction as it comes from a circular convolution, periodic boundary conditions for $\rho^T$ guarantee that the targets' mass is conserved, i.e. $\left( \int_\Omega \rho^T\xt \, \mathrm{d} \mathbf{x} \right)_t = 0$. Initial conditions are set as
\begin{subequations}
\begin{align}
    \rho^H(\mathbf{x},0) = \rho^H_0(\mathbf{x}), \\
    \rho^T(\mathbf{x},0) = \rho^T_0(\mathbf{x}), 
\end{align}
\end{subequations}
such that $\int_\Omega \rho^i_0(\mathbf{x}) \, \mathrm{d} \mathbf{x} = M^i$, $i=T,H$. 

\subsection{Macroscopic control design}\label{sec:macro_control_design}
We divide the control design into two sequential steps. First, we determine the number of herders $N^H$ required to steer the targets towards the desired density. Next, we design the control velocity field $\mathbf{u}$ to ensure that \eqref{eq:continuum_problem_statement} is satisfied. This approach leverages the results of  \cite{maffettone2024leaderfollower}. 

\subsubsection{Choice of $N^H$}
To determine the appropriate number of herders $N^H$, we draw the feasibility analysis carried out in \cite{maffettone2024leaderfollower}. 

\begin{definition}
    We say that problem \eqref{eq:continuum_problem_statement}-\eqref{eq:LF_continuum_model} admits a feasible solution if, given a targets' mass $0<M^T<1$, there exists some herders' density $\Bar{\rho}^H(\mathbf{x})$ fulfilling the following two conditions:
    \begin{enumerate}
        \item $\Bar{\rho}^H(\mathbf{x})\geq 0, \quad \forall \mathbf{x} \in \Omega$,
        \item $\int_\Omega \Bar{\rho}^H(\mathbf{x}) \, \mathrm{d} \mathbf{x} = M^H = 1 - M^T$,
    \end{enumerate}
    and such that the desired targets' density $\Bar{\rho}^T$ is a solution of \eqref{eq:targets_continuum} with $\mathbf{v}^{TH}\xt = \Bar{\mathbf{v}}^{TH}(\mathbf{x}) \coloneq \left(\mathbf{f} * \Bar{\rho}^H\right)(\mathbf{x})$.
\end{definition}
Hence, given the desired target density profile $\Bar{\rho}^T$, we assume it to be a solution of \eqref{eq:targets_continuum} and derive an expression for $\Bar{\mathbf{v}}^{TH}$ as
\begin{equation}
    \Bar{\mathbf{v}}^{TH} (\mathbf{x}) = D\frac{\nabla \Bar{\rho}^T(\mathbf{x})}{\Bar{\rho}^T(\mathbf{x})}.
\end{equation}
Recalling that $\Bar{\mathbf{v}}^{TH} = \mathbf{f}*\Bar{\rho}^H$, we can obtain $\Bar{\rho}^H$ by its deconvolution with $\mathbf{f}$ \cite{wing1991primer}, leading to

\begin{equation} \label{eq:rho_bar_L}
    \Bar{\rho}^H(\mathbf{x}) = H(\mathbf{x}) + A,
\end{equation}
where $H$ is the deconvolution of $\mathbf{v}^{TH}$ and $A$ is an arbitrary constant\footnote{Since a closed form for $\mathbf{f}$ was not found, the deconvolution can only be carried out numerically (see Appendix \ref{sec:numerical_deconvolution} for more details).}. We remark that the deconvolution operation does not necessarily return non-negative functions.

\begin{proposition}
    Given $\Bar{\rho}^H$ as defined in \eqref{eq:rho_bar_L} and setting $A = \min_\mathbf{x} H(\mathbf{x})$, the quantity
    \begin{equation} \label{eq:minimum_herders_mass}
        \widehat{M}^H = \int_\Omega \Bar{\rho}^H(\mathbf{x}) \, \mathrm{d} \mathbf{x}
\end{equation}
represents the minimum herders' mass required to make the problem feasible.
\end{proposition}
With this choice of $A$ we ensure that $\Bar{\rho}^H$ is non-negative and has the minimum possible integral.
\begin{figure}
    \centering
    \includegraphics[width=0.72\linewidth]{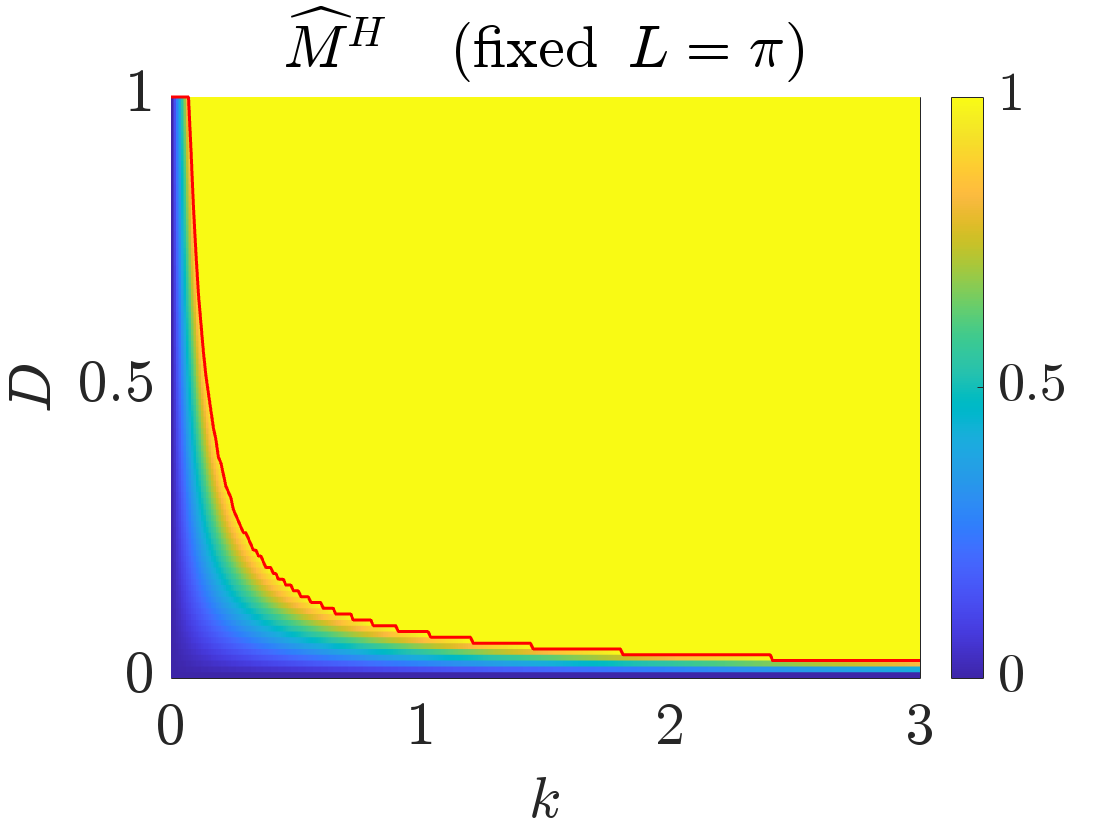}
    \caption{Feasibility plot showing the minimum amount of herders' mass $\widehat{M}^H$ varying $\mathbf{k}$ and $D$ of the Von-Mises distribution \eqref{eq:von_mises} with fixed kernel length $L=\pi$. The red line denotes the curve where $\widehat{M}^H$ becomes greater than $1$. $\widehat{M}^H$ has been saturated to 1 for visualization purposes.}
    \label{fig:M_hat_L}
\end{figure}
For a graphical interpretation of \eqref{eq:minimum_herders_mass}, we provide a feasibility plot in Fig. \ref{fig:M_hat_L}. Specifically, we set $\Bar{\rho}^T$ to be the Von-Mises distribution in \eqref{eq:von_mises} and, after fixing the characteristic length scale of the interaction kernel at $L=\pi$, we analyzed how $\widehat{M}^H$ changes when varying the diffusion coefficient $D$, and the concentration coefficients of the desired targets’ distribution $k$ (assuming $k_1 = k_2 = k$); the yellow area in Fig. \ref{fig:M_hat_L} being the region in which the problem becomes unfeasible.

We are then able to recover the minimum amount of herders from $\widehat{M}^H$ as
\begin{equation}
    N^H = N^T \frac{\widehat{M}^H}{1-\widehat{M}^H}.
\end{equation}

\begin{figure} 
    \centering
    \includegraphics[width=1\linewidth]{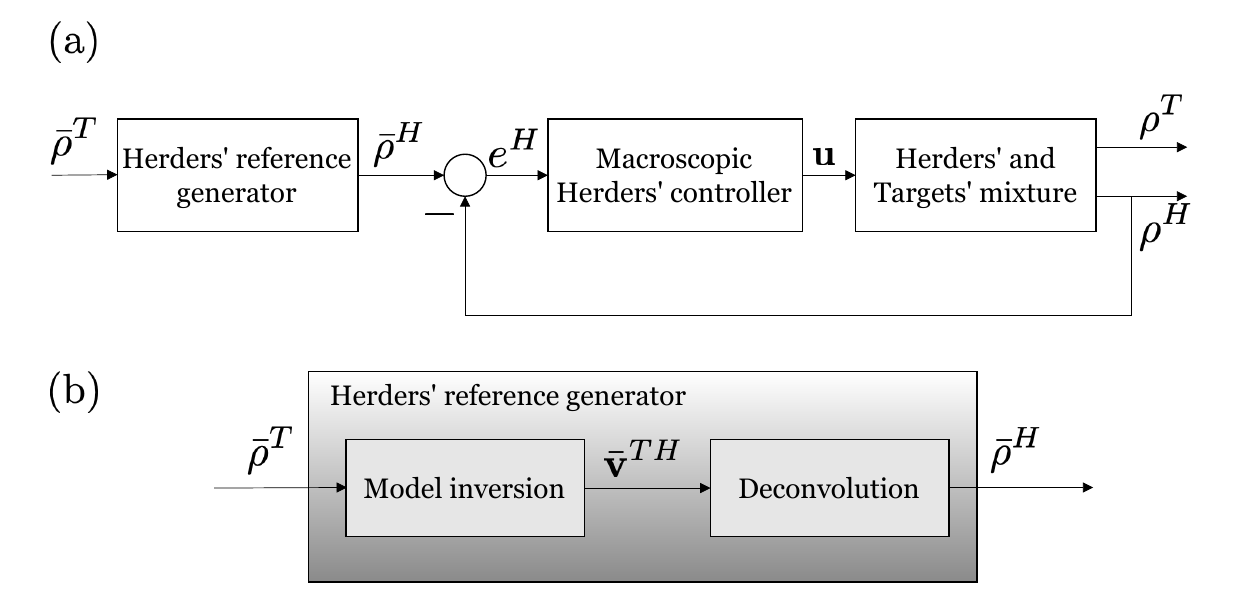}%
    \caption{(a) Macroscopic feed-forward control scheme. (b) Detail of the feasibility analysis step.}
    \label{fig:FF_scheme}
\end{figure}

\subsubsection{Design of $\mathbf{u}$}
We fix $\mathbf{u}$ in \eqref{eq:herders_continuum} using the the feed-forward strategy that is discussed in \cite{maffettone2024leaderfollower} (see Section VI for the one-dimensional case), whose scheme is depicted in Fig. \ref{fig:FF_scheme}. Notably, this solution is a feed-forward with respect to the targets' dynamics, but incorporates feedback for the herders' dynamics. Specifically, we choose $\mathbf{u}$ so that herders asymptotically displace according to $\Bar{\rho}^H$ in \eqref{eq:rho_bar_L}. We then prove that, once the herders achieve $\Bar{\rho}^H$, the targets globally asymptotically achieve $\bar{\rho}^T$.

\paragraph{herders' control design} Given the desired herders' density profile $\Bar{\rho}^H$, we want to design $\mathbf{u}$ in order to drive $\rho^H$ towards this desired profile. We define the herders' density error as
\begin{equation}
    e^H\xt = \Bar{\rho}^H(\mathbf{x}) - \rho^H\xt.
\end{equation}

\begin{theorem}[\textbf{Herders' global exponential convergence}]
Let $\mathbf{u}$ be chosen such that
\begin{equation} \label{eq:herders_control_law}
    \nabla \cdot \left[ \rho^H\xt \mathbf{u}\xt \right] = -K_H e^H\xt
\end{equation}
with $K_H>0$. Under this condition, the herders' error globally exponentially converges to zero. Additionally, such a choice of $\mathbf{u}$ is periodic.
\end{theorem}

For the proof of convergence see Sec. X.C of \cite{maffettone2024leaderfollower}.

\paragraph{Targets' stability analysis}
We demonstrate that, under suitable conditions, the targets density $\rho^T$ also converges to the desired profile $\Bar{\rho}^T$. We define the targets' error as
\begin{equation} \label{eq:targets_error}
    e^T\xt = \Bar{\rho}^T(\mathbf{x}) - \rho^T\xt.
\end{equation}
The error dynamics is given by
\begin{equation} \label{eq:targets_error_dynamics}
    e_t^T \xt = \nabla \cdot \left[ \rho^T\xt \mathbf{v}^{TH} \xt \right] - D \nabla^2 \rho^T \xt,
\end{equation} 
where $e^T$ is periodic on $\partial \Omega$ with initial conditions $e^T(\mathbf{x},0) = \Bar{\rho}^T(\mathbf{x}) - \rho_0^T(\mathbf{x})$ $\forall \mathbf{x} \in \Omega$.

Let us define
\begin{equation}
    G(\mathbf{x}) = \nabla \cdot \frac{\nabla \Bar{\rho}^T(\mathbf{x})}{\Bar{\rho}^T(\mathbf{x})}.
\end{equation}
\begin{theorem}[\textbf{Targets' global exponential stability}]
    If $\| G(\mathbf{x}) \|_\infty < 2$, the error dynamics \eqref{eq:targets_error_dynamics} globally exponentially converges to 0 in $\mathcal{L}^2(\Omega)$, that is
    \begin{equation}
        \left\| e^T(\cdot,t) \right\|_2^2 \leq \left\| e^T(\cdot,0) \right\|_2^2 \exp\{ -K^{ff}t \},
    \end{equation}
    with $K^{ff} = D(2 - \| G \|_\infty)$.
\end{theorem}
The proof can be found in Theorem 5 of \cite{maffettone2024leaderfollower}.

\subsection{Discretization}
Here, we discretize the macroscopic, continuum control action $\mathbf{u}$ given in \eqref{eq:herders_control_law} into a set of discrete control inputs $\mathbf{u}_i(t)$, $i=1,\dots,N^H$ to be deployed on the herders.

As in \cite{maffettone2024mixed}, in order to uniquely recover $\mathbf{u}$ from \eqref{eq:herders_control_law}, we add an extra condition on the curl of $(\rho^H\mathbf{u})$. This is necessary because \eqref{eq:herders_control_law} constitutes a scalar relation for the unknown vector field $\mathbf{u}$.
Hence, we define
\begin{equation}
    \mathbf{w}\xt \coloneq \rho^H \xt \mathbf{u} \xt
\end{equation}
and close the problem by adding an extra condition on the curl of $\mathbf{w}$
\begin{equation} \label{eq:closed_problem}
    \begin{cases}
        \nabla \cdot \mathbf{w} \xt = -K_H e^H \xt, \\
        \nabla \times \mathbf{w} = 0,
    \end{cases}
\end{equation}
where $\mathbf{w}$ is periodic on $\partial \Omega$. Note that problem \eqref{eq:closed_problem} is purely spatial, as it involves no time derivatives. Since $\mathbf{w}$ is irrotational and $\Omega$ is simply connected, we can express it as the gradient of a scalar potential, that is $\mathbf{w} = -\nabla \varphi$. Substituting this into \eqref{eq:closed_problem}, we obtain the Poisson problem
\begin{equation} \label{eq:poisson_problem}
    \nabla^2 \varphi \xt = -K_H e^H\xt,
\end{equation}
which can be solved using Fourier series expansion, following the approach described in \cite{maffettone2024mixed}. We expand $\varphi$ in $\Omega$ as 
\begin{equation} \label{eq:phi_fourier}
    \varphi (\mathbf{x}) = \sum_{\mathbb{m}\in \mathbb{Z}^2} \gamma_\mathbb{m} \mathrm{e}^{\mathrm{j} \mathbf{m}\cdot \mathbf{x}} + C
\end{equation}
where $\mathbb{m}$ is a multi-index, $\mathbf{m}$ is the row vector associated with it, $\gamma_\mathbb{m}$ is the $\mathbb{m}$-th Fourier coefficient, $\mathrm{j}$ is the imaginary unit and $C$ is an arbitrary constant. We write the Laplacian as
\begin{equation} \label{eq:phi_laplacian}
    \nabla^2 \varphi(\mathbf{x}) = \sum_{\mathbb{m}\in \mathbb{Z}^2} \gamma_\mathbb{m} \| \mathbf{m} \|_2^2 \mathrm{e}^{\mathrm{j} \mathbf{m}\cdot \mathbf{x}}.
\end{equation}

Similarly, we express the Fourier series of $e^H$ as 
\begin{equation} \label{eq:error_fourier}
    e^H (\mathbf{x}) = \sum_{\mathbb{m}\in \mathbb{Z}^2} c_\mathbb{m} \mathrm{e}^{\mathrm{j} \mathbf{m}\cdot \mathbf{x}}
\end{equation}
where
\begin{equation} 
    c_\mathbb{m} = \frac{1}{(2\pi)^2} \int_\Omega e^H (\mathbf{x}) \mathrm{e}^{- \mathrm{j} \mathbf{m} \cdot \mathbf{x}} \, \mathrm{d} \mathbf{x}
\end{equation}
are the Fourier coefficients of $e^H$. Therefore, to satisfy \eqref{eq:poisson_problem}, the Fourier coefficients in \eqref{eq:phi_laplacian} and \eqref{eq:error_fourier} must be equal, that is
\begin{equation}
    \gamma_\mathbb{m} = -K_H\frac{c_\mathbb{m}}{\| \mathbf{m} \|_2^2}.
\end{equation}
Once we obtain $\varphi$, we get $\mathbf{w}=-\nabla \varphi$ and, consequently,
\begin{equation}
    \mathbf{u} \xt = \frac{\mathbf{w}\xt}{\rho^H \xt}.
\end{equation}
Note that $\mathbf{u}$ is well-defined only when $\rho^H>0$. This assumption is reasonable since the density is ultimately estimated using an estimation kernel of our choice (for instance choosing a Gaussian kernel estimator ensures the strict positivity of the density). Moreover, $\mathbf{u}$ will be sampled at the agents' positions, where $\rho^H>0$.

We can then spatially sample the macroscopic control velocity field at the agents' positions:
\begin{equation}
    \mathbf{u}_i(t) = \mathbf{u}(\mathbf{H}_i,t), \quad i=1,\dots,N^H.
\end{equation}
From an implementation viewpoint, the Fourier series representing $\varphi$ is truncated to a sufficiently large number of terms in the summation \eqref{eq:phi_fourier}.

\section{VALIDATION} \label{sec:validation}
In this section, we validate our control strategy through simulations and experiments in the Robotarium \cite{robotarium}.

\subsection{Numerical validation}\label{sec:num_validation}
We select the goal region as a circle centered at $\mathbf{p} = \mathbf{0}$ with radius $r^*=\pi/2$, resulting in $\mu=\nu=0$ and $k_1 = k_2 = 6/\pi$. We set $D=0.01$, $L=\pi$, $K_H=10$ and $N^T=720$. Using our feasibility result, we determine that the number of herders required to make the problem feasible is $N^H = 280$.

\begin{figure}
    \centering
    \includegraphics[width=1\linewidth]{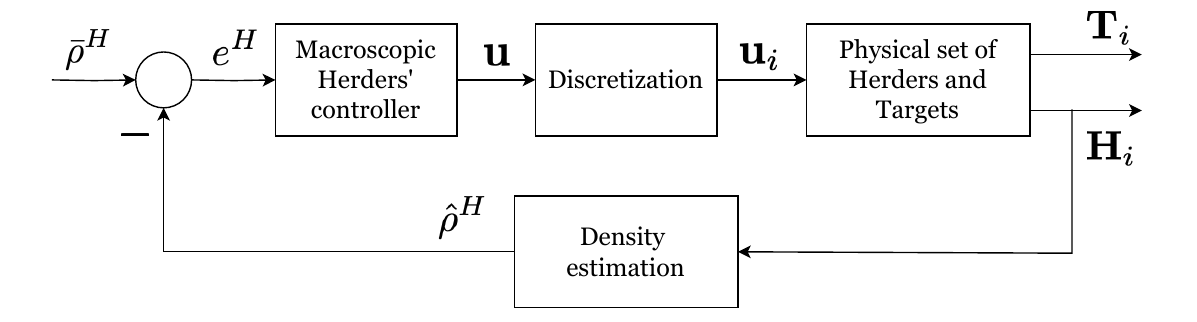}
    \caption{Implementation scheme: adaptation of the macroscopic controller described in Sec. \ref{sec:macro_control_design} to cope with a discrete set of herders and targets; the ``Discretization" and ``Density estimation" blocks work across the micro and macro scale.}
    \label{fig:implementation_scheme}
\end{figure}

For implementation at the microscopic agents' level of the macroscopic controller designed in Sec. \ref{sec:macro_control_design} and schematized in Fig. \ref{fig:FF_scheme}, we  adopt the schematic depicted in Fig. \ref{fig:implementation_scheme}. Specifically, an estimate of the herder's density, say $\hat{\rho}^H$, is obtained using a Gaussian kernel estimation method, based on the positions of the microscopic agents. To this aim, we adapted the MATLAB  function \textit{ksdensity} for periodic domains, where multivariate Gaussian kernels are located at each agents' position and summed to obtain the estimated density. The bandwidth for each kernel is heuristically set to 0.4 (this parameter influences the resulting agents' displacement; optimizing its choice is the subject of ongoing work). The agents' dynamics \eqref{eq:agents} is integrated with the Euler-Maruyama method, fixing a time step of $\Delta t = 0.01$ over a time horizon of $T = 200$.
The herders start from a rectangular lattice formation, whereas the initial positions of the targets are randomly drawn from a uniform distribution.

\begin{figure}
    \centering
    \includegraphics[width=1\linewidth]{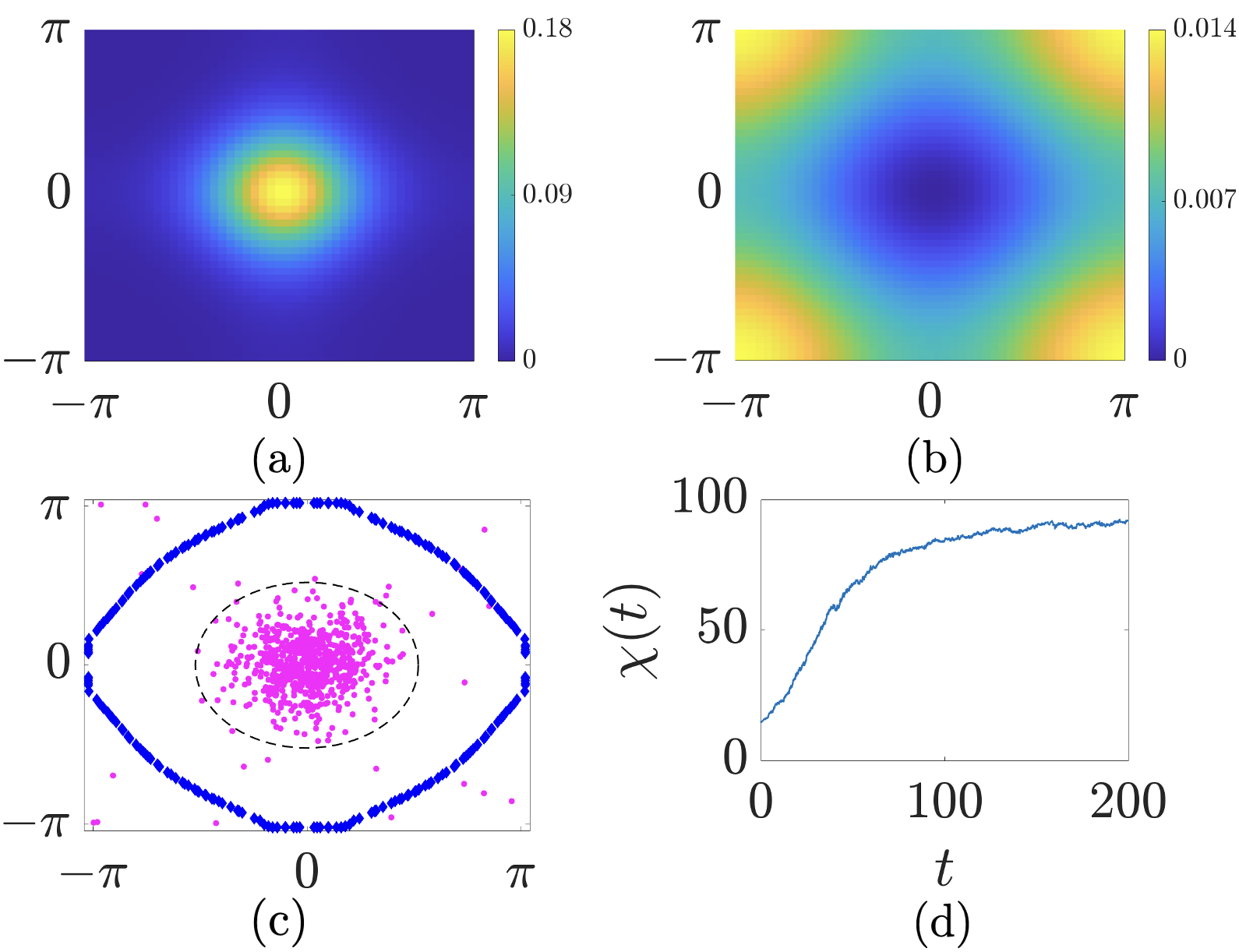}
    \caption{(a) Desired targets' density $\bar{\rho}^T$. (b) Desired herders' density $\bar{\rho}^H$. (c) Final agents' configuration: the goal region is represented by the dashed circle, blue diamonds represent the herders, magenta dots are the targets. (d) Percentage of targets inside the goal region.}
    \label{fig:numerical_validation}
\end{figure}

To assess performance, we consider the percentage of target agents inside the goal region at time $t$, defined as
\begin{equation}
    \chi (t) = \frac{N^T_{in} (t)}{N^T} \cdot 100
\end{equation}
where $N^T_{in}$ represents the number of targets inside the goal region. In Fig. \ref{fig:numerical_validation}a and \ref{fig:numerical_validation}b, we show the desired targets' concentration $\bar{\rho}^T$ and the corresponding herders' distribution $\bar{\rho}^H$ obtained with the procedure discussed in Sec. \ref{sec:macro_control_design}. The final configuration of the agents, and the time evolution of $\chi$ are depicted in Fig. \ref{fig:numerical_validation}c and \ref{fig:numerical_validation}d, with $92\%$ of the targets successfully confined within the goal region. For the simulation's video, we refer to our Github repository {\tt\small{\url{https://bit.ly/3AsllNY}}}.

\subsection{Experimental validation}
We further validate our control strategy using the Robotarium, an open access platform developed at GeorgiaTech \cite{robotarium, pickem2017robotarium}. The Robotarium is a remotely accessible facility for swarm robotics research, consisting of a $3.2 \, \mathrm{m} \times 2 \, \mathrm{m}$ arena where differential drive robots are used to test control algorithms. Up to 20 robots can be utilized for experiments; each robot measures $11 \, \mathrm{cm} \times 10 \, \mathrm{cm} \times 7 \, \mathrm{cm}$, with a maximum linear speed of $20 \, \mathrm{cm/s}$ and a maximum rotational speed of $3.6 \, \mathrm{rad/s}$. The Robotarium implements its own collision avoidance protocols, and, if needed, transforms velocity inputs for omni-directional point masses into suitable accelerations for the robots.

We designate the available 20 differential-drive robots to be herders, while targets are simulated via software. Note that the total number of agents we can consider is limited by the total number of robots provided by the platform. Therefore, similar to \cite{maffettone2024mixed}, we simulate $N^T = 80$ virtual targets modeled as in\eqref{eq:targets_discrete} with a time step $\Delta t = 0.01$ for a time horizon $T=70$, setting $D=0.01$, $L=\pi$, and $K_H=10$. 

To adapt the periodicity assumption of the domain to the experimental setup, we define a smaller domain $\Omega_A = [-1, 1]^2$ within the arena and scale the robots' positions relative to this fictitious domain. The virtual targets, instead, are allowed to cross the boundaries since they are simulated agents. The initial configuration is set as in numerical simulations -- see Sec. \ref{sec:num_validation}.

We set $r^*_A = 0.5$, to scale the previously chosen $r^*=\pi/2$ with respect to the fictitious domain $\Omega_A$ and adjust the concentration coefficient of $\bar{\rho}^T$ accordingly to $k=6/\pi$.

\begin{figure} 
    \centering
    \subfloat[$t = 0$, $\chi = 11\%$]{%
        \includegraphics[width=0.3\textwidth]{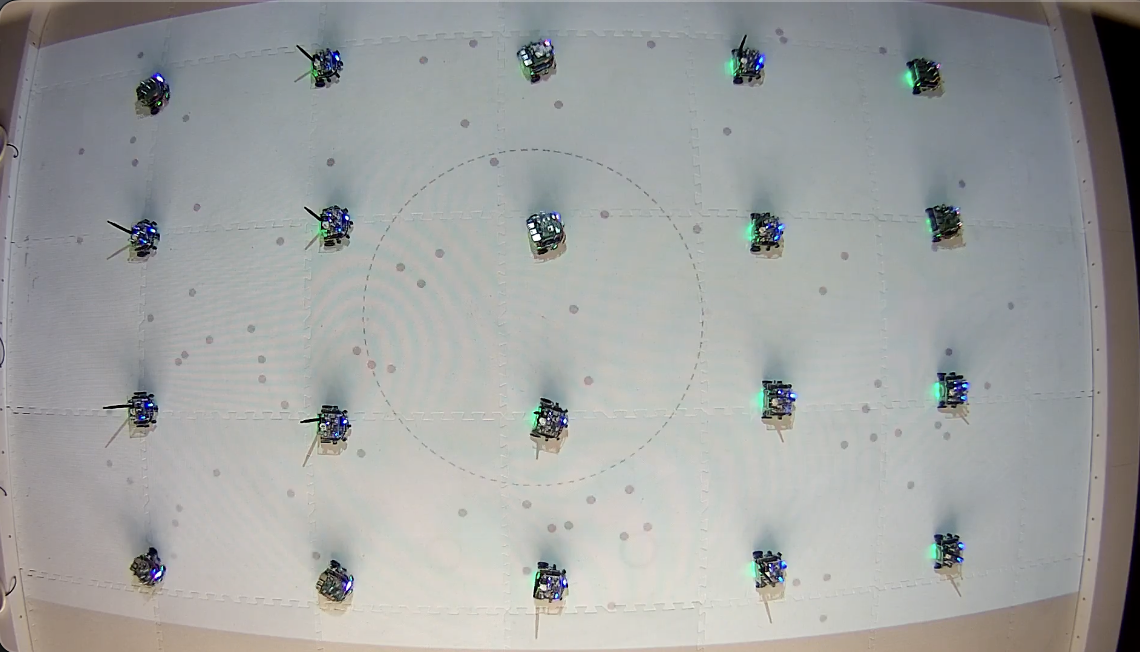}%
        \label{fig:robotarium_initial_config}%
        }%
    \hfill%
    \subfloat[$t = 70$, $\chi = 100\%$]{%
        \includegraphics[width=0.3\textwidth]{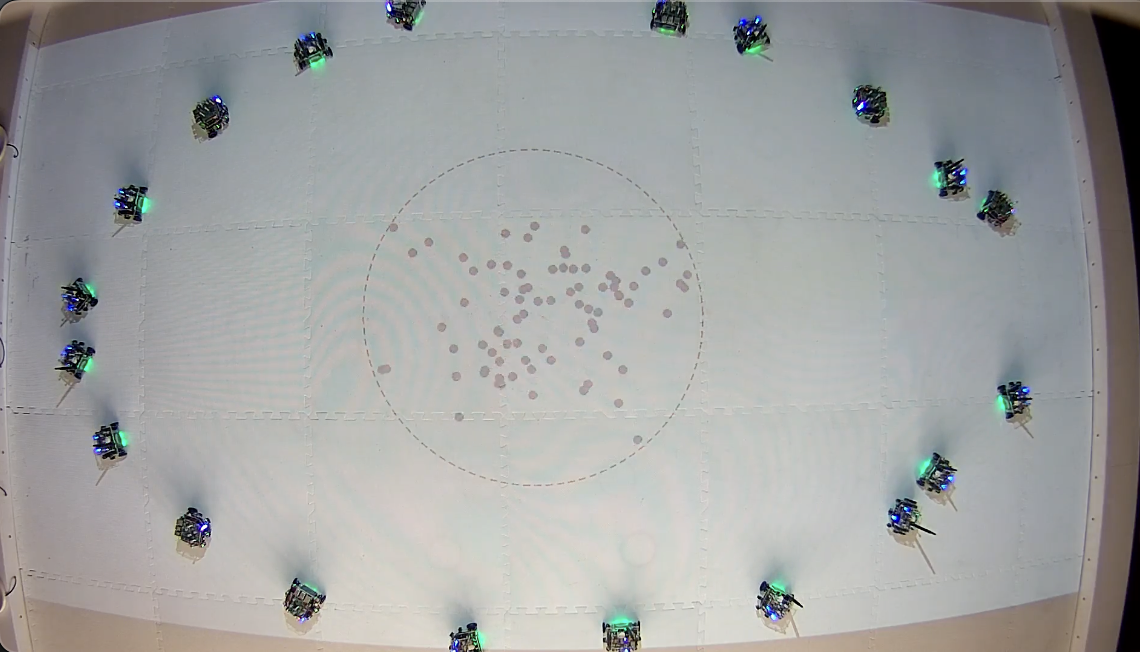}%
        \label{fig:robotarium_final_config}%
        }%
    \caption{(a) Initial and (b) final configuration of the swarm in Robotarium.}
    \label{fig:experimental_validation}
\end{figure}

In Fig. \ref{fig:experimental_validation}, we present the initial and final configurations of the robots and virtual agents. Notably, starting with only 11\% of the targets inside the goal region, we successfully corralled 100\% of the population at steady state. The final configuration closely resembles the ``encirclement" (or caging) solution to the shepherding problem, which is often observed in natural systems and highlighted in the literature as a typical solution for this type of problems \cite{varava2017herding,van2023steering,van2024reactive}. For a video of the experiment, we refer the reader to our GitHub repository at {\tt\small{\url{https://bit.ly/3AsllNY}}}.

\section{CONCLUSIONS} \label{sec:conclusions}
In this work, we propose a continuification-based approach to solve the large-scale shepherding control problem. By building on the theoretical results from density control in leader-follower frameworks \cite{maffettone2024leaderfollower}, we extended these principles to a large-scale multi-agent context. The novelty of our approach lies in the systematic adaptation of continuification techniques to shepherding, including the formulation of a method to compute the minimal number of herders required to achieve containment of a large population of targets within a goal region; a problem closely related to that of studying their ``herdability'' as recently defined in \cite{lama2024shepherding}.

The novelty of our results also lies in bridging the gap between theory and practical implementation, providing a tangible demonstration of how macroscopic control strategies can be effectively used in mixed-agent systems.
Our numerical simulations and experiments in Robotarium demonstrate the feasibility of applying a continuification-based approach to real-world scenarios, underscoring the scalability and effectiveness of our control strategy.

Future work will investigate alternative solutions for controlling a large number of targets with a limited number of herders, possibly utilizing models involving coupled ODEs and PDEs, such as in \cite{albi2016invisible}. 

\appendix
\subsection{Numerical deconvolution} \label{sec:numerical_deconvolution}
Since the analytical expression of the periodic kernel $\mathbf{f}$ is not available (see \cite{boldini2024stigmergy, maffettone2024mixed}), the deconvolution in \eqref{eq:rho_bar_L} must be performed numerically. To achieve this, we define a square mesh of $M \times M$ evenly displaced cells in $\Omega$ with fixed step $h$ and define the grid points $\mathbf{q}_i$, $i=1,\dots,M^2$. The convolution integral $\bar{\mathbf{v}}^{TH} = \mathbf{f}*\bar{\rho}^H$ evaluated at $\mathbf{q}_i$ can be approximated using the summation
\begin{multline} \label{eq:approx_conv}
    \bar{\mathbf{v}}^{TH} (\mathbf{q}_i) = \sum_{j=1}^M \sum_{k=1}^M h^2 w_{jk} \dots \\ \dots \mathbf{f} \left(\{ \mathbf{q}_i,(y_{1,j},y_{2,k}) \} \right) \bar{\rho}^H(y_{1,j},y_{2,k}), \quad i=1,\dots,M^2,
\end{multline}
where $w_{jk}$ are the quadrature weights of the trapezoidal rule for integration and $y_{1,j}$, $y_{2,k}$ are the first and second components of the $j$-th and $k$-th grid points. By recasting the elements of the summation in a matrix $\mathbf{F}$, we can write $\bar{\mathbf{v}}^{TH}\approx \mathbf{F} \bar{\rho}^H$. Hence, by pseudo-inversion of matrix $\mathbf{F}$ we obtain
\begin{equation}
    \bar{\rho}^H(\mathbf{x}) = \mathbf{F}^\dag (\mathbf{x}) \bar{\mathbf{v}}^{TH} (\mathbf{x}) + B
\end{equation}
where $\mathbf{F}^\dag$ is the left-inverse of $\mathbf{F}$ and $B$ is an arbitrary constant. Since $\mathbf{f}$ is assumed to be an odd function with respect to any possible direction, and due to the linearity of the convolution operator, the convolution integral is defined up to a constant, i.e. $\bar{\mathbf{v}}^{TH}(\mathbf{x}) = \left( \mathbf{f}*\bar{\rho}^H \right)(\mathbf{x}) = \left( \mathbf{f}* (\bar{\rho}^H + B) \right)(\mathbf{x})$. 
This leads to \eqref{eq:rho_bar_L}.

\bibliographystyle{IEEEtran}

\begin{thebibliography}{10}
\providecommand{\url}[1]{#1}
\csname url@samestyle\endcsname
\providecommand{\newblock}{\relax}
\providecommand{\bibinfo}[2]{#2}
\providecommand{\BIBentrySTDinterwordspacing}{\spaceskip=0pt\relax}
\providecommand{\BIBentryALTinterwordstretchfactor}{4}
\providecommand{\BIBentryALTinterwordspacing}{\spaceskip=\fontdimen2\font plus
\BIBentryALTinterwordstretchfactor\fontdimen3\font minus
  \fontdimen4\font\relax}
\providecommand{\BIBforeignlanguage}[2]{{%
\expandafter\ifx\csname l@#1\endcsname\relax
\typeout{** WARNING: IEEEtran.bst: No hyphenation pattern has been}%
\typeout{** loaded for the language `#1'. Using the pattern for}%
\typeout{** the default language instead.}%
\else
\language=\csname l@#1\endcsname
\fi
#2}}
\providecommand{\BIBdecl}{\relax}
\BIBdecl

\bibitem{maffettone2022continuification}
G.~C. Maffettone, A.~Boldini, M.~di~Bernardo, and M.~Porfiri,
  ``Continuification control of large-scale multiagent systems in a ring,''
  \emph{IEEE Control Systems Letters}, vol.~7, pp. 841--846, 2022.

\bibitem{maffettone2024leaderfollower}
G.~C. Maffettone, A.~Boldini, M.~Porfiri, and M.~di~Bernardo, ``Leader-follower
  density control of spatial dynamics in large-scale multi-agent systems,''
  \emph{arXiv preprint arXiv:2406.01804}, 2024.

\bibitem{nikitin2021continuation}
D.~Nikitin, C.~Canudas-de Wit, and P.~Frasca, ``A continuation method for
  large-scale modeling and control: from odes to pde, a round trip,''
  \emph{IEEE Transactions on Automatic Control}, vol.~67, no.~10, pp.
  5118--5133, 2021.

\bibitem{bongini2017optimal}
M.~Bongini and G.~Buttazzo, ``Optimal control problems in transport dynamics,''
  \emph{Mathematical Models and Methods in Applied Sciences}, vol.~27, no.~03,
  pp. 427--451, 2017.

\bibitem{fornasier2014mean}
M.~Fornasier and F.~Solombrino, ``Mean-field optimal control,'' \emph{ESAIM:
  Control, Optimisation and Calculus of Variations}, vol.~20, no.~4, pp.
  1123--1152, 2014.

\bibitem{d2023controlling}
R.~M. D’Souza, M.~di~Bernardo, and Y.-Y. Liu, ``Controlling complex networks
  with complex nodes,'' \emph{Nature Reviews Physics}, vol.~5, no.~4, pp.
  250--262, 2023.

\bibitem{lama2024shepherding}
A.~Lama and M.~di~Bernardo, ``Shepherding and herdability in complex multiagent
  systems,'' \emph{Physical Review Research}, vol.~6, p. L032012, 7 2024.

\bibitem{long2020comprehensive}
N.~K. Long, K.~Sammut, D.~Sgarioto, M.~Garratt, and H.~A. Abbass, ``A
  comprehensive review of shepherding as a bio-inspired swarm-robotics guidance
  approach,'' \emph{IEEE Transactions on Emerging Topics in Computational
  Intelligence}, vol.~4, no.~4, pp. 523--537, 2020.

\bibitem{jyh2004shepherding}
J.-M. Lien, O.~B. Bayazit, R.~T. Sowell, S.~Rodriguez, and N.~M. Amato,
  ``Shepherding behaviors,'' in \emph{Proceedings of IEEE International
  Conference on Robotics and Automation. ICRA'04.}, vol.~4.\hskip 1em plus
  0.5em minus 0.4em\relax IEEE, 2004, pp. 4159--4164.

\bibitem{zahugi2013oil}
E.~M.~H. Zahugi, M.~M. Shanta, and T.~Prasad, ``Oil spill cleaning up using
  swarm of robots,'' in \emph{Proceedings of the Second International
  Conference on Advances in Computing and Information Technology (ACITY) July
  13-15, 2012, Chennai, India-Volume 3}.\hskip 1em plus 0.5em minus 0.4em\relax
  Springer, 2013, pp. 215--224.

\bibitem{yuan2023multi}
Z.~Yuan, T.~Zheng, M.~Nayyar, A.~R. Wagner, H.~Lin, and M.~Zhu,
  ``Multi-robot-assisted human crowd control for emergency evacuation: A
  stabilization approach,'' in \emph{Proceedings of the 2023 American Control
  Conference (ACC)}.\hskip 1em plus 0.5em minus 0.4em\relax IEEE, 2023, pp.
  4051--4056.

\bibitem{maffettone2024mixed}
G.~C. Maffettone, L.~Liguori, E.~Palermo, M.~di~Bernardo, and M.~Porfiri,
  ``Mixed reality environment and high-dimensional continuification control for
  swarm robotics,'' \emph{IEEE Transactions on Control Systems Technology},
  vol. 32, no. 6, pp. 2484--2491, 2024.

\bibitem{robotarium}
S.~Wilson, P.~Glotfelter, L.~Wang, S.~Mayya, G.~Notomista, M.~Mote, and
  M.~Egerstedt, ``The robotarium: Globally impactful opportunities, challenges,
  and lessons learned in remote-access, distributed control of multirobot
  systems,'' \emph{IEEE Control Systems Magazine}, vol.~40, no.~1, pp. 26--44,
  2020.

\bibitem{ko2020asymptotic}
D.~Ko and E.~Zuazua, ``Asymptotic behavior and control of a “guidance by
  repulsion” model,'' \emph{Mathematical Models and Methods in Applied
  Sciences}, vol.~30, no.~04, pp. 765--804, 2020.

\bibitem{boldini2024stigmergy}
A.~Boldini, M.~Civitella, and M.~Porfiri, ``Stigmergy: from mathematical
  modelling to control,'' \emph{Royal Society Open Science}, vol.~11, no.~9, p.
  240845, 2024.

\bibitem{wing1991primer}
G.~M. Wing, \emph{A primer on integral equations of the first kind: the problem
  of deconvolution and unfolding}.\hskip 1em plus 0.5em minus 0.4em\relax SIAM,
  1991.

\bibitem{pickem2017robotarium}
D.~Pickem, P.~Glotfelter, L.~Wang, M.~Mote, A.~Ames, E.~Feron, and
  M.~Egerstedt, ``The robotarium: A remotely accessible swarm robotics research
  testbed,'' in \emph{2017 IEEE International Conference on Robotics and
  Automation (ICRA)}.\hskip 1em plus 0.5em minus 0.4em\relax IEEE, 2017, pp.
  1699--1706.

\bibitem{varava2017herding}
A.~Varava, K.~Hang, D.~Kragic, and F.~T. Pokorny, ``Herding by caging: a
  topological approach towards guiding moving agents via mobile robots.'' in
  \emph{Robotics: Science and Systems}, 2017, pp. 1--9.

\bibitem{van2023steering}
S.~Van~Havermaet, P.~Simoens, T.~Landgraf, and Y.~Khaluf, ``Steering herds away
  from dangers in dynamic environments,'' \emph{Royal Society Open Science},
  vol.~10, no.~5, p. 230015, 2023.

\bibitem{van2024reactive}
S.~Van~Havermaet, Y.~Khaluf, and P.~Simoens, ``Reactive shepherding along a
  dynamic path,'' \emph{Scientific Reports}, vol.~14, no.~1, p. 14915, 2024.

\bibitem{albi2016invisible}
G.~Albi, M.~Bongini, E.~Cristiani, and D.~Kalise, ``Invisible control of
  self-organizing agents leaving unknown environments,'' \emph{SIAM Journal on
  Applied Mathematics}, vol.~76, no.~4, pp. 1683--1710, 2016.

\end{thebibliography}

\end{document}